\documentclass{ifacconf}

\usepackage{graphicx}      
\usepackage{natbib}        
\usepackage{amsmath}       
\usepackage{amssymb}       
\usepackage{xcolor}
\usepackage{url}
\usepackage{algpseudocode}
\usepackage{algorithmicx,algorithm}
\usepackage{diagbox} 
\usepackage{adjustbox}

\newcounter{WSQcomment}

\newcounter{FMYcomment}

\begin{document}
\begin{frontmatter}

\title{Chance-Constrained Neural MPC under Uncontrollable Agents via  Sequential Convex Programming\thanksref{footnoteinfo}} 

\thanks[footnoteinfo]{This work was supported by the National Natural Science Foundation of China (62173226,62061136004).}

\author[First,Second]{Shuqi Wang} 
\author[First]{Mingyang Feng} 
\author[First]{Yu Chen} 
\author[First,Second]{Yue Gao}
\author[First]{Xiang Yin}

\address[First]{School of Automation and Intelligent Sensing, Shanghai Jiao Tong University, Shanghai, China (e-mail:\{wangshuqi,fmy-135214,yuchen26,yuegao,yinxiang\}@sjtu.edu.cn).}
\address[Second]{Shanghai Innovation Institute, Shanghai, China}

\begin{abstract}                %
This work investigates the challenge of ensuring safety guarantees in the presence of uncontrollable agents, whose behaviors are stochastic and depend on both their own and the system's states.
We present a neural model predictive control (MPC) framework that predicts the trajectory of the uncontrollable agent using a predictor learned from offline data. 
To provide formal probabilistic guarantees on prediction errors despite policy-induced distribution shifts, we propose a region-wise robust conformal prediction scheme to construct time-dependent uncertainty bounds, which are integrated into the MPC formulation.
To solve the resulting non-convex, discontinuous optimization problem, we propose a two-loop iterative sequential convex programming algorithm. The inner loop solves convexified subproblems with fixed error bounds, while the outer loop refines these bounds based on updated control sequences. We establish convergence guarantees and analyze the optimality of the algorithm.
We illustrate our method with an autonomous driving scenario involving interactive pedestrians.
Experimental results demonstrate that our approach achieves superior safety and efficiency compared to baseline methods, with success rates exceeding 99.5\% while maintaining higher average speeds in multi-pedestrian scenarios.
\end{abstract}
\end{frontmatter}


\section{Introduction}
Safety is a fundamental requirement for autonomous systems operating in dynamic, uncertain environments, such as autonomous driving and human-robot interactions. To achieve safe control in real-world scenarios, it is crucial to account for the presence of \emph{uncontrollable agents} in the environment~(\cite{sadigh2016planning}). A particular challenge arises when these environments involve \emph{coupled agents}, in which the behavior of an uncontrollable agent depends on the state of the controllable system~(\cite{fisac2019hierarchical, wang2024s4tp}). For example, in autonomous driving, a pedestrian may slow down or change direction in response to an approaching vehicle. This coupling significantly complicates both prediction and planning, as the evolution of each agent is no longer independent.

Another fundamental challenge comes from the unknown nature of the internal decision-making processes of uncontrollable agents. Classical robust control approaches often assume worst-case disturbances, which can lead to overly conservative behaviors and poor scalability, particularly in interactive environments. Stochastic control frameworks, such as chance-constrained MPC, model agent behaviors as known stochastic processes to provide probabilistic safety guarantees~(\cite{blackmore2011chance,cc-scp}). However, the true distributions governing these behaviors are typically unknown, rendering the underlying optimization problems analytically intractable.

More recently, data-driven approaches have emerged, in which neural networks are trained on real-world trajectory data to predict the future behavior of other agents~(\cite{rhinehart2019precog, casas2020implicit, ejaz2023trust}). 
Parallel to these advancements in prediction, \cite{neural-mpc} presented a Neural MPC framework that integrates neural networks into the MPC pipeline to model complex system dynamics. While powerful, such frameworks typically do not provide well-calibrated uncertainty estimates, making them unsuitable for direct use in safety-critical control pipelines. 
In response to this gap, there has been growing interest in integrating \emph{distribution-free uncertainty quantification} methods such as \emph{conformal prediction} into planning and control. Conformal prediction allows one to wrap any black-box predictor with statistical guarantees on the prediction error, under mild assumptions~(\cite{angelopoulos2021gentle}), making it promising for building safe control frameworks that leverage data-driven models~(\cite{sun2024conformal, lindemann2023}). 

\begin{figure*}[htbp]
  \centering
  \includegraphics[width=0.9\textwidth]{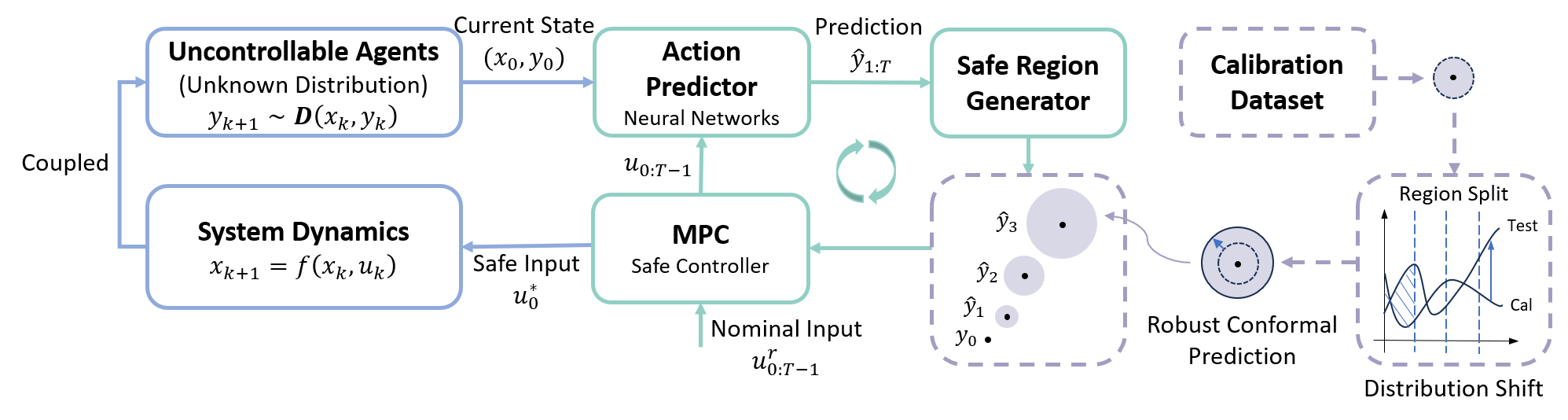}
  \caption{Overview of the proposed safe control framework. In interactive environments, predicting uncontrollable agents while planning often creates a logical deadlock: the ego vehicle's policy adjustments change the pedestrian's future distribution, which in turn invalidates the original policy. To break this deadlock, our Action Predictor explicitly conditions the pedestrian's future nominal trajectory ($\hat{y}_{1:T}$) on the vehicle's planned control sequence ($u_{0:T-1}$). Simultaneously, the Safe Region Generator constructs a corresponding robust error bound via Robust Conformal Prediction, which rigorously compensates for the policy-induced distribution shifts between offline calibration and online testing. By integrating these dynamically coupled predictions and robust bounds into a two-loop SCP architecture, our framework solves for a control input ($u_{0}^{*}$) that strictly satisfies probabilistic safety guarantees in real time.}
  \label{fig:framework}
\end{figure*}

However, existing CP-based control frameworks often face  two major issues in interactive environments. First, the ego system's control policy during online testing inherently differs from the calibration data collection policy, inducing a \emph{policy-induced distribution shift} that invalidates standard exchangeability assumptions. Second, predicting uncontrollable agents while simultaneously planning creates a fundamental \emph{logical deadlock}: the ego system's policy adjustments change the uncontrollable agents' future distributions, which in turn invalidates the predictive safety bounds used to compute that very policy. Standard approaches lack the mechanism to systematically decouple this circular dependency, making it difficult to simultaneously guarantee strict safety and optimal performance.

In this work, we propose a robust safe control framework for systems with coupled uncontrollable agents. We model the agent's behavior as a state-dependent distribution learned via neural networks. To address distribution shifts, we develop a region-wise robust conformal prediction scheme to construct mathematically guaranteed error bounds. These bounds are integrated into an MPC formulation, which we solve using a novel two-loop iterative sequential convex programming (SCP) algorithm. We provide a complete convergence and optimality analysis of the algorithm.

Our overall contributions are summarized as follows:
\begin{itemize}
\item We consider a general model for coupled dynamics with uncertainty, where the trajectory prediction network accounts for potential future control signals and captures the interdependence between the controllable system and uncontrollable agents.\medskip
\item We propose a region-wise robust conformal prediction framework that provides formal probabilistic safety guarantees under policy-induced distribution shifts, explicitly accounting for the discrepancy between offline calibration and online execution.\medskip
\item We develop a two-loop iterative SCP algorithm to solve the resulting non-convex, discontinuous optimization problem. We provide a convergence analysis, proving that the algorithm terminates in finite steps, yielding KKT-optimal / $\epsilon_{\text{tol}}$-optimal solutions under normal conditions, or a strictly feasible fallback solution to guarantee safety.\medskip
\item We validate the method in a high-fidelity simulator modeling real-world social interactions, demonstrating both the safety and efficiency of the approach in a realistic setting.
\end{itemize}

\textbf{Related Works.}\quad 
There are several works on behavior predictions and uncertainty quantification for safe control. 
In \cite{ivanov2020} and \cite{waite2025state}, the authors explore how neural networks propagate errors over time in the form of reachable sets. These studies primarily focus on quantifying noise uncertainty in the perception module. In contrast, our work assumes perfect state observation but aims to quantify and control the uncertainty in the behavior of uncontrollable agents.
In the context of using conformal prediction (CP) for safe control in the presence of uncontrollable agents, \cite{lindemann2023} proposed a framework that integrates CP into MPC. However, their method assumes that the future behavior of dynamic agents follows a stationary distribution independent of the system's state. 

In \cite{dixit2023adaptive}, the authors introduced an adaptive conformal prediction framework that recalibrates uncertainty online throughout each trajectory, improving robustness under distribution shifts. However, this approach assumes that at each step, the behavior of uncontrollable agents depends only on the last state of the controllable system, neglecting the agents' own state. Crucially, their online recalibration mechanism operates in an \textit{a posteriori} manner, introducing an inherent temporal latency. 
This framework relies on realized empirical errors from preceding horizons to iteratively update future uncertainty bounds. Consequently, it exhibits delayed responsiveness to abrupt, policy-induced distribution shifts. While it achieves marginal coverage over a long-term horizon, this latency inherently compromises instantaneous safety during highly dynamic interactions.

\emph{Notations: }
We denote $\mathbb{R}$, $\mathbb{N}$, and $\mathbb{R}^n$ as the sets of real numbers, natural numbers, and $n$-dimensional real vectors, respectively. 
For any vector $v \in \mathbb{R}^n$, we use $|v|$ and $\|v\|$ to denote its $\ell_1$-norm and Euclidean norm, respectively.  For a finite set $D$, we denote $|D|$ as its cardinality. 
We use $C^1$ to denote the class of continuously differentiable functions.

\section{Problem Formulation and Preliminaries}

\subsection{System Models}
In this work, we consider a scenario where a controlled system operates in an environment involving uncontrollable agents.
For simplicity, our theoretical developments focus on a single uncontrollable agent; however, the results can be naturally extended to multiple uncontrollable agents, as demonstrated in our experiments despite the associated scalability challenges.
We assume that the dynamics of the controlled system are known, while the behavior (i.e., internal control law) of the uncontrollable agent remains unknown.
Our objective is to design a control law for the controlled system that ensures safety in the presence of such interacting uncontrollable agents.

\textbf{Controlled System:} 
We model the controllable agent as a discrete-time  nonlinear system 
\begin{equation}\label{eq:controllable_dynamics}
x_{k+1} = f(x_k, u_k), 
\end{equation}
where 
$x_k \in \mathcal{X} \subseteq \mathbb{R}^{n_x}$ denotes the system state,
$u_k \in \mathcal{U} \subseteq\mathbb{R}^{n_u}$ is the control input, and 
$f:\mathbb{R}^{n_x}\times \mathbb{R}^{n_u}\to \mathbb{R}^{n_x}$  
represents the dynamics of the system. We assume both $\mathcal{X}$ and $\mathcal{U} $ are convex sets.
The system dynamic $f$  is \emph{perfectly known} and its state $x_k$ at each   instant is fully observable.

\textbf{Uncontrollable Agents:} 
The dynamics of the uncontrollable agent are modeled as an \emph{unknown distribution} 
\begin{equation}
 y_{k+1} \,\sim \, \mathcal{D}(x_k, y_k), 
\label{eq:uncontrollable_dynamics}
\end{equation}
where \( y_k \in \mathcal{Y} \subseteq \mathbb{R}^{n_y} \) represents the state of the uncontrollable agent. $\mathcal{D}(x_k, y_k)$ governs the next state of the uncontrollable agent, and depends on both the controllable system state \( x_k \) and the uncontrollable agent's state \( y_k \).
Such a state-dependent and unknown distribution naturally arises in many practical scenarios. For instance, in autonomous driving, the internal decision-making process of a pedestrian is unknown, stochastic, and influenced by the vehicle's state, such as its speed and distance to the pedestrian.

\subsection{Problem Formulation}
A control sequence from time step $k_1$ to $k_2$ is written as $\mathbf{u}_{k_1:k_2} := (u_{k_1},\ldots,u_{k_2})$. Similarly, we define $\mathbf{x}_{k_1:k_2}$ and $\mathbf{y}_{k_1:k_2}$ for state sequences. For brevity, we use the simplified notation $\mathbf{u}$, $\mathbf{x}$, and $\mathbf{y}$ to denote the complete sequences $\mathbf{u}_{0:T-1}$, $\mathbf{x}_{1:T}$, and $\mathbf{y}_{1:T}$, respectively. 
The safety of the system is characterized by a Lipschitz continuous function $c:\mathbb{R}^{n_x}\times\mathbb{R}^{n_y}\to \mathbb{R}$ with Lipschitz constant $L$ 
such that $c(x_k,y_k)\geq 0$ is considered safe. We assume that the initial states satisfy   $c(x_0,y_0)\geq 0$.

Our objective is to design a feedback control law that provides formal safety guarantees for the system. 
To achieve this, we employ a model predictive control (MPC) framework.
Specifically, at each time step, we solve an open-loop  chance-constrained optimal control problem (CCOCP) with built-in safety guarantees. 
Such an optimization problem will be solved iteratively in a receding horizon fashion.

\begin{prob}[\textbf{CCOCP}]Find a  control input sequence $ \mathbf{u}^*$ on the time horizon $T$
that minimizes a cost function $J$ convex on $\mathbf{u}$ while ensuring safety with a probability $1-\gamma$ for each time step. That is 
\begin{equation}
\min_{\mathbf{u} }  J(\mathbf{x},\mathbf{u})
\end{equation}
subject to, for all $k = 1, \dots, T$:
\begin{align*}
&\mathbb{P} \left[ c(x_{k}, y_{k}) \geq 0  \right] \geq 1-\gamma, \\
    & x_k, y_k \text{ satisfy the dynamics  } \eqref{eq:controllable_dynamics} \text{ and }\eqref{eq:uncontrollable_dynamics}, \\
     & u_k \in \mathcal{U}, x_k \in \mathcal{X}.
\end{align*}
\label{prb:ccocp}
\end{prob}
In the above problem formulation, note that, given the initial state $(x_0,y_0)$ and the control sequence $\mathbf{u}_{0:T-2}$, the state sequence of the controlled system $\mathbf{x}_{0:T-1}$ can be uniquely determined by \eqref{eq:controllable_dynamics}. Based on this, the state sequence of the uncontrollable agent $\mathbf{y}_{1:T}$, 
which are random variables, can also be derived through \eqref{eq:uncontrollable_dynamics}. 

\subsection{Data Collection and Trajectory Predictor}
\label{sec:data_collection}
To solve the chance-constrained optimization problem, we predict a nominal trajectory $\hat{\mathbf{y}}_{1:T}$, and then expand this trajectory into a sequence of reachable regions with a high-confidence bound, transforming the problem into a deterministic optimization problem.

\textbf{Trajectory Predictor.} \quad 
To obtain a nominal trajectory for the uncontrollable agent, we assume the availability of a trajectory predictor, denoted by $\Omega$. The input to the predictor is of the form  $X= (x_0,y_0,\mathbf{u}_{0:T-2})$, which includes the current state and the designed inputs. The output of the predictor  is a $T$-step forecast of the uncontrollable agent, denoted by $\hat{Y} = \Omega(X) = (\hat{y}_1, \ldots, \hat{y}_T)$.
To train the trajectory predictor, 
we assume a dataset $D$ is  available, where each data sample is of the form $(X, Y)$ sampled from a data-generating distribution $\mathcal{P}$. Such data can be extracted from sliding windows of long trajectories. For later purposes, we split the dataset  $D$  into training and calibration datasets $D_{\text{train}}$ and $D_{\text{cal}}$, respectively, and assume that the predictor $\Omega$  is trained from $D_{\text{train}}$. A specific example of a trajectory predictor is recurrent neural networks (RNNs), such as gated recurrent units (GRUs) (\cite{chung2014empiricalevaluationgatedrecurrent}), which have demonstrated good performance in our experiments.

\subsection{Conformal Prediction}
To ensure probabilistic safety guarantees under uncertainty, we employ \emph{conformal prediction} (\cite{shafer2008tutorial}).
Specifically, we recall the following result, which validates the coverage region of predictors with a probability equal to or exceeding a given confidence level.

\begin{lem}[\cite{tibshirani2019conformal}, Lemma 1]\upshape
\label{lem:cp}
Let $R^{(0)},\\ \ldots, R^{(K)}$ be $K+1$ exchangeable real-valued random variables.\footnote{Exchangeability is a slightly weaker form of independent and identically distributed (i.i.d.) random variables.} Given a failure probability tolerance $\gamma \in [0,1]$,
it holds that $\operatorname{Prob}\left(R^{(0)} \leq \bar{R}(\gamma) \right) \geq 1-\gamma$, where 
$    \bar{R}(\gamma) = \text { Quantile }_{1-\gamma}\left(R^{(1)}, \ldots, R^{(K)}, \infty\right).$
\end{lem}
The variable $R^{(i)}$ is usually referred to as the \textit{nonconformity score}. 
In supervised learning, it is often defined as the difference between the model output and the ground-truth.
In our work, we will use a computationally more tractable variant called the \emph{split conformal prediction}; see, e.g., \cite{Papadopoulos08}.

\section{Robust Conformal Prediction under Distribution Shift}

In closed-loop control, the trajectory predictor evaluates states that are dynamically induced by the current MPC policy, which inherently differs from the offline data collection policy. This difference induces a standard covariate shift: the marginal distribution of the input states changes ($\mathcal{P}_{\text{test}}(X) \neq \mathcal{P}_{\text{cal}}(X)$), while the environment's underlying conditional distribution of the prediction error given a specific state ($\mathcal{L}(R_k \mid X)$) remains invariant. 
To formally bound the impact of this policy-induced distribution shift, our approach is to localize the problem and formulate a region-wise robust conformal prediction scheme. 


Formally, let $\mathbb{X}$ denote the input space. We define a measurable partition $\mathbb{P} := \{\mathbb{X}_1, \dots, \mathbb{X}_M\}$ such that $\mathbb{X} = \dot{\bigcup}_{p=1}^M \mathbb{X}_p$. Given the split conformal calibration dataset $D_{\text{cal}} = \{(X^{(i)}, Y^{(i)})\}_{i=1}^{N_{\text{cal}}}$, we construct the region-specific calibration subset as:
\[
D_{\mathbb{X}_p}^{XY} := \{(X^{(i)}, Y^{(i)}) \in D_{\text{cal}} \mid X^{(i)} \in \mathbb{X}_p\}.
\]
Let $S_k: \mathbb{X} \times \mathbb{Y} \to \mathbb{R}_+$ be the nonconformity score at time step $k$. In this work, we define the prediction error $R_k := S_k(X, Y) = \|\hat{y}_k(X) - y_k\|_2$. The corresponding empirical score set for region $\mathbb{X}_p$ is given by:
\[
D_{\mathbb{X}_p}^{R_k} := \{R_k^{(i)} \mid (X^{(i)}, Y^{(i)}) \in D_{\mathbb{X}_p}^{XY}\}.
\]

Our approach relies on the conformal prediction error bound, computed for each partitioned region, to quantify uncertainty. However, a fixed closed-loop policy in test can cause the actual distribution of prediction errors to deviate from the calibration data, even within the same region. To account for such potential distribution shifts, we assume that the error distribution remains relatively stable when the input state changes only slightly, such as within a small partitioned region. Mathematically, this assumption is characterized by the local Lipschitz regularity of the distribution distance with respect to the input state.


\begin{assum}[Local Lipschitz Regularity]
\label{assum:lipschitz}
The underlying unknown conditional score distribution $\mathcal{L}(R_k \mid X=x)$ is locally Lipschitz continuous with respect to the input state in the Total Variation (TV) metric. That is, within each local region $\mathbb{X}_p$, there exists a finite local Lipschitz constant $L_R > 0$ such that for all $x, x' \in \mathbb{X}_p$:
\[
d_{\mathrm{TV}}\big(\mathcal{L}(R_k \mid X=x), \mathcal{L}(R_k \mid X=x')\big) \le L_R \|x - x'\|,
\]
where $d_{\mathrm{TV}}(\cdot,\cdot)$ denotes the TV distance between two probability distributions; see, e.g., \cite{TV_distance}.
\end{assum}


Based on Assumption~\ref{assum:lipschitz}, for a fixed partition size, one can leverage the regional diameter $\Delta_p$ alongside the Lipschitz constant $L_R$ to quantify the distribution shift. Consequently, this allows for the derivation of a robust prediction error bound that remains valid under policy-induced distribution shifts.
Conversely, if the objective is to achieve a predefined error tolerance, the parameter $L_R$ can be employed to inversely determine the required precision of the partition grid. 
Fortunately, the qualitative relationship between the robust error bound and the partition size can be characterized by the following fundamental result.
\begin{lem}
(Robust Conformal Prediction, \cite{barber2023conformal})
\label{lem:robust_cp}
If the Total Variation distance between the test distribution and the calibration distribution is bounded by $\delta \in [0, 1)$, then computing the $(1-\gamma+\delta)$-quantile on the calibration nonconformity scores guarantees a predictive coverage of at least $1-\gamma$ on the test distribution.
\end{lem}

We define the regional diameter $\Delta_p := \sup_{x, x' \in \mathbb{X}_p} \|x - x'\|$. 
Building upon the above assumption and lemma, the following theorem establishes how this regional diameter $\Delta_p$ allows us to compute a robustified empirical quantile that strictly guarantees the desired predictive coverage.
\begin{thm}[Region-Wise Robust Coverage]
\label{thm:region_coverage}
Suppose Assumption~\ref{assum:lipschitz} holds. Let $\delta_p := L_R \Delta_p$. If $\delta_p < \gamma$, and we define the robustified region-wise prediction bound as the shifted empirical quantile:
\[
\bar{R}^{\mathrm{rob}}_{k,\mathbb{X}_p} := \widehat{Q}_{1 - \gamma + \delta_p}\big(D_{\mathbb{X}_p}^{R_k}\big),
\]
then the predictive coverage on the test distribution, conditioned on $X \in \mathbb{X}_p$, strictly satisfies:
\[
\mathbb{P}_{\mathrm{test}} \big(R_k \le \bar{R}^{\mathrm{rob}}_{k,\mathbb{X}_p} \mid X \in \mathbb{X}_p\big) \ge 1 - \gamma, \quad \forall k \in \{1, \dots, T\}.
\]
\end{thm}
\begin{pf}
Let $\mathcal{P}_{\text{test}}(X \mid X \in \mathbb{X}_p)$ and $\mathcal{P}_{\text{cal}}(X \mid X \in \mathbb{X}_p)$ denote the conditional distributions of the input states within the region $\mathbb{X}_p$ under the closed-loop test policy and the offline calibration dataset, respectively. The overall distributions of the nonconformity scores within this region are essentially mixture distributions marginalized over these state distributions:
\begin{align*}
    \mathcal{L}_{\text{test}}(R_k) &= \int_{\mathbb{X}_p} \mathcal{L}(R_k \mid X=x) \, \mathrm{d}\mathcal{P}_{\text{test}}(x \mid X \in \mathbb{X}_p), \\
    \mathcal{L}_{\text{cal}}(R_k) &= \int_{\mathbb{X}_p} \mathcal{L}(R_k \mid X=x') \, \mathrm{d}\mathcal{P}_{\text{cal}}(x' \mid X \in \mathbb{X}_p).
\end{align*}

By the convexity of the TV distance, the distance between these two mixture distributions is bounded by the expected distance between their underlying conditional components: $d_{\mathrm{TV}} \big(\mathcal{L}_{\text{test}}(R_k), \mathcal{L}_{\text{cal}}(R_k)\big)\le \int_{\mathbb{X}_p} \int_{\mathbb{X}_p} d_{\mathrm{TV}} \big(\mathcal{L}(R_k \mid X=x), \mathcal{L}(R_k \mid X=x')\big) \, \mathrm{d}\mathcal{P}_{\text{test}}(x) \, \mathrm{d}\mathcal{P}_{\text{cal}}(x').$

According to Assumption~\ref{assum:lipschitz}, for any $x, x' \in \mathbb{X}_p$, the distance between the conditional distributions is bounded by $L_R \|x - x'\|$. Furthermore, since both states are strictly confined within the partitioned region $\mathbb{X}_p$, their maximum spatial distance is universally bounded by the region's diameter $\Delta_p$. Thus, we have:
\begin{align*}
    &d_{\mathrm{TV}} \big(\mathcal{L}_{\text{test}}(R_k), \mathcal{L}_{\text{cal}}(R_k)\big) \\
    & \le \int_{\mathbb{X}_p} \int_{\mathbb{X}_p} L_R \|x - x'\| \, \mathrm{d}\mathcal{P}_{\text{test}}(x) \, \mathrm{d}\mathcal{P}_{\text{cal}}(x') \\
    &\le \int_{\mathbb{X}_p} \int_{\mathbb{X}_p} L_R \Delta_p \, \mathrm{d}\mathcal{P}_{\text{test}}(x) \, \mathrm{d}\mathcal{P}_{\text{cal}}(x') \\
    &= L_R \Delta_p = \delta_p.
\end{align*}

This establishes that the maximum distribution shift between the test and calibration scores within the region $\mathbb{X}_p$ is upper-bounded by $\delta_p$. Applying Lemma~\ref{lem:robust_cp}, computing the shifted quantile $\widehat{Q}_{1 - \gamma + \delta_p}$ on the regional calibration set inherently compensates for this shift, thereby guaranteeing a valid lower bound predictive coverage of $1-\gamma$ on the test distribution. \hfill $\blacksquare$
\end{pf}
\begin{rem}(Discussion on the Local Lipschitz Regularity Assumption)
\label{rem:lipschitz_discussion}
In Assumption~\ref{assum:lipschitz}, we assume that the conditional distribution of the prediction error is locally Lipschitz continuous with respect to the input state under the TV metric. This assumption is realistic for most continuous physical interactions. For instance, when navigating alongside a human in an open corridor, the uncontrollable agent's distancing or lateral avoidance behavior typically shifts smoothly in response to gradual changes in the ego system's state. In practice, the Lipschitz constant $L_R$ can be empirically estimated from existing operational data of the uncontrollable agents or derived analytically from their physical and kinematic constraints. 

Furthermore, we acknowledge that in certain specific regions, such as discrete decision boundaries where a pedestrian abruptly switches from ``yielding" to ``crossing" at an intersection, the distribution shift can be sharp. In such cases, one could conservatively take the supremum of the local constants as a global parameter to guarantee strict safety, though this inevitably leads to more conservative planning. A more practical and less conservative alternative would be to employ region-adaptive constants $L_{R,p}$ tailored to each specific partition $\mathbb{X}_p$. Nevertheless, for the sake of notational simplicity and theoretical clarity, we utilize a uniform global Lipschitz constant $L_R$ in our current formulation. 
\end{rem}


\section{Deterministic Problem Reformulation and Algorithm} 

\subsection{Deterministic Reformulation}

Recall that in Problem~\ref{prb:ccocp}, we require the probabilistic safety constraint $\mathbb{P}[c(x_k, y_k) \ge 0] \ge 1-\gamma$.
Given the Lipschitz continuity of the safety function $c$ with constant $L$, we have $|c(x_k, y_k) - c(x_k, \hat{y}_k)| \le L \|y_k - \hat{y}_k\|_2 = L R_k$.

By utilizing the robustified error bound derived in Section III, we have $\mathbb{P}_{\text{test}}(R_k \le \bar{R}^{\mathrm{rob}}_{k,\mathbb{X}_p}) \ge 1 - \gamma$.
Therefore, enforcing the deterministic constraint $c(x_k, \hat{y}_k) \ge L \bar{R}^{\mathrm{rob}}_{k,\mathbb{X}_p}$ is strictly sufficient to guarantee the original constraint.
This leads to the following Deterministic Optimal Control Problem (DOCP), which is a reformulation
of Problem~\ref{prb:ccocp}:


\begin{prob}[\textbf{DOCP}]
\label{prb:docp}
Given current state $(x_0,y_0)$ with $c(x_0,y_0)\geq 0$,
find a control input sequence $ \mathbf{u}^*$ that minimizes the cost while ensuring deterministic safety, i.e.,
\begin{equation}
\min_{\mathbf{u} } J(\mathbf{x},\mathbf{u})
\end{equation}
subject to, for all $k = 1, \dots, T$:
\begin{align*}
&x_{k+1}=f(x_{k}, u_{k}),\\
&\hat{\mathbf{y}}_{1:T} = \Omega(x_0,y_0,\mathbf{u}_{0:T-2}),\\
&c(x_k,\hat{y}_{k} )\geq L\bar{R}^{\mathrm{rob}}_{k,\mathbb{X}_p}, \\
     & u_k \in \mathcal{U}, x_k \in \mathcal{X}.
\end{align*}
\end{prob}

However, the above DOCP remains non-convex due to the nonlinear system dynamics $f$, the overly complex high-dimensional neural networks $\Omega$, and coupled safety constraints $c$. Standard convex optimization approaches such as \cite{dixit2023adaptive} would require $c \circ f$, $c \circ \Omega$ to be concave functions of $\mathbf{u}$, 
these conditions are overly restrictive in practice.

One may consider adopting some existing non-linear solvers or algorithms, such as sequential convex programming (SCP, \cite{svanberg1987method}) techniques to solve the optimization problem. 
However, this faces additional challenges. Specifically, the error bounds $\bar{R}_{k,\mathbb{X}_p}$ are inherently discontinuous functions of the control sequence $\mathbf{u}$, as they depend on discrete region assignments in the partitioned state space. This discontinuity violates the $C^1$ smoothness requirement necessary for standard non-linear programming solvers. Furthermore, the complex, black-box nature of the neural network predictor $\Omega$, with respect to the control variable $\mathbf{u}$, is also unacceptable.

To address these challenges, we propose a two-loop iterative SCP algorithm that decouples the prediction error bound updates from the approximate convex optimization iterations, enabling effective handling of both the non-convex dynamics and the discontinuous error bounds. 

\subsection{Two-Loop Iterative Algorithms}

Here, we propose  a two-loop SCP algorithm to solve the  DOCP. 
Specifically, the \emph{inner loop} (Algorithm~\ref{alg:scp_inner}) solves the convex subproblem for fixed error bounds, while the \emph{outer loop} (Algorithm~\ref{alg:scp_outer}) refines the error bounds based on updated optimal control sequences solved by the inner loop. 
We will show that, through finite steps, the Algorithm~\ref{alg:scp_outer} will terminate and return one of these kinds of solutions: (1) Feasible (line 6-8). If the updated error bounds shift such that the current solution violates the safety constraints, the solution becomes infeasible and cannot serve as a valid initial point for the subsequent iteration. To strictly guarantee safety, the algorithm aborts the current optimization loop and falls back to return the previous valid solution $\mathbf{u}^{*(\ell-1)}$, which is strictly a feasible solution within its corresponding feasible region, but it is not guaranteed to be a KKT optimal solution.
(2) KKT optimal (line 9-11). If the updated error bounds remain unchanged or expand (i.e., triggering the shortcut), the algorithm directly returns the current feasible and KKT optimal  $\mathbf{u}^{*(\ell)}$. We will later establish it in Theorem~\ref{thm:outer_convergence}. 
(3) $\epsilon_\text{tol}$-optimal (line 13). If neither early termination condition is triggered, the algorithm monotonically descends the cost function until it slows down. Although this returned solution may not be exactly KKT optimal for its corresponding feasible region, it is an $\epsilon$-optimal solution.

The key property of the returned solution is strict closed-loop safety. Subject to this hard requirement, the algorithm systematically seeks a high-quality solution for the DOCP (Problem~\ref{prb:docp}). Next, we provide rigorous mathematical definitions of the inner loop and outer loop, and then analyze the properties mentioned above.

\begin{algorithm}[t]
\caption{SCP (for Fixed Error Bounds)}
\label{alg:scp_inner}
\hspace*{0.02in} {\bf Input:} 
Initial states $x_t, y_t$; Fixed error bounds $\bar{R}$; initial control $\mathbf{u}^{(0)}$; trust region $\Delta_0$\\
\hspace*{0.02in} {\bf Output:} 
Optimal control sequence $\mathbf{u}^*$
\begin{algorithmic}[1]
\State Initialize $j \gets 0$, $\Delta_j \gets \Delta_0$
\Repeat
    \State Linearize $\tilde{x}_k^{(j)},\tilde{c}^{(j)}$ at $\mathbf{u}^{(j)}$, as in  Eq.~\eqref{eq:lin_x}\eqref{eq:linearized_constraint} 
  \State Solve LOCP \eqref{prb:locp} to obtain $\mathbf{u}^{(j+1)}$
  \State Update trust region $\Delta_{j+1}$ adaptively based on the linear approximation accuracy
  \State $j \gets j + 1$
\Until{convergence or max iterations}
\State \Return $\mathbf{u}^* = \mathbf{u}^{(j)}$
\end{algorithmic}
\end{algorithm}

\begin{algorithm}[t]
\caption{MPC with Error Bounds Update}
\label{alg:scp_outer}
\hspace*{0.02in} {\bf Input:} Current states $x_t, y_t$; tolerance $\epsilon_{\text{tol}}$; Initial conservative control sequence $\mathbf{u}_0$ \\
\hspace*{0.02in} {\bf Output:} Control sequence $\mathbf{u}^*_{t:t+T-1}$ (denoted as $\mathbf{u}^*$)
\begin{algorithmic}[1]
\State Initialize $\ell \gets 0$, $\mathbf{u}^{*(0)} \gets \mathbf{u}_0$
\Repeat
    \State Update $\mathbf{u}^{*} \gets \mathbf{u}^{*(\ell)}$, $\ell \gets \ell + 1$
    \State Retrieve $\bar{R}^{(\ell)}$ using $\mathbf{u}^{*(\ell-1)}$ by Eq.~\eqref{eq:construct_bound}
  \State$\mathbf{u}^{*(\ell)} \gets$ SCP $(x_t, y_t, \bar{R}^{(\ell)}, \mathbf{u}^{*(\ell-1)})$
    \State Retrieve $\bar{R}^{(\ell+1)}$ using $\mathbf{u}^{*(\ell)}$ by  Eq.~\eqref{eq:construct_bound}
  \If{$g^{(\ell+1)}(\mathbf{u}^{*(\ell)}) > 0$}
        \State Reject \textit{infeasible} $\mathbf{u}^{*(\ell)}$,  \Return $\mathbf{u}^{*(\ell-1)}$
    \EndIf
    \If {$\bar{R}^{(\ell+1)} \ge \bar{R}^{(\ell)}$}
      \State \Return \textit{shortcut} $\mathbf{u}^{*(\ell)}$
    \EndIf

\Until{$J({\textbf{u}}^{*(\ell-1)}) - J({\textbf{u}}^{*(\ell)})\leq \epsilon_{\text{tol}}$}
\State \Return $\mathbf{u}^{*}$
\end{algorithmic}
\end{algorithm}

    \textbf{Outer Loop (Error Bounds Update):} For the $\ell$-th outer loop, refine the error bounds 
    \begin{equation}
    \label{eq:construct_bound}
        \bar{R}_k^{(\ell)} := \bar{R}^{\mathrm{rob}}_{k,\mathbb{X}_p} ~\text{where}~  (x_0, y_0, \mathbf{u}^{*(\ell-1)}_{0:T-2}) \in \mathbb{X}_p
    \end{equation}
    based on the updated control sequence $\mathbf{u}^{*(\ell-1)}$ solved in the ($\ell$-1)-th outer loop. Denote $\{\bar{R}_k^{(\ell)}\}_{k=1}^T$ as $\bar{R}^{(\ell)}$. Then, we can construct the \textbf{$\ell$-th DOCP}  parameterized by $\bar{R}^{(\ell)}$. Next, one can use SCP to compute the next solution $\mathbf{u}^{*(\ell)}$ within several inner loops.

    \textbf{Inner Loop (SCP):} For fixed error bounds 
    $\bar{R}$, we iteratively solve convexified subproblems with linearized constraints until convergence at $\mathbf{u}^{*}$. The working principle of SCP is successively linearizing the costs and nonconvex constraints, seeking a solution of the original problem through a sequence of convex problems. Given the solution $\mathbf{u}^{(j)}$ from the convexified problem at iteration $j$, the convex approximation of (DOCP) at the current iteration $j+1$ is described next.

First, we linearize the controllable agent's state trajectory with respect to the control sequence. Since the dynamics $x_{k+1} = f(x_k, u_k)$ in \eqref{eq:controllable_dynamics} are known, applying the chain rule recursively yields that each state $x_k$ can be expressed as an affine function of $\mathbf{u}$ around the reference trajectory $\mathbf{u}^{(j)}$:
\begin{equation}
\label{eq:lin_x}
    \tilde{x}_k^{(j)}(\mathbf{u}) = x_k^{(j)} + \nabla_{\mathbf{u}} x_k(\mathbf{u}^{(j)})^\top (\mathbf{u} - \mathbf{u}^{(j)}),
\end{equation}
where $\nabla_{\mathbf{u}} x_k$ is computed recursively via the chain rule: $\nabla_{\mathbf{u}} x_{k+1} = \nabla_x f(x_k, u_k) \cdot \nabla_{\mathbf{u}} x_k + \nabla_u f(x_k, u_k)$ with initial condition $\nabla_{\mathbf{u}} x_0 = 0$. This linearization enables us to transform the state constraint $x_k \in \mathcal{X}$ into a convex constraint $\tilde{x}_k^{(j)}(\mathbf{u}) \in \mathcal{X}$. 

For the nonconvex safety constraints $c(x_k,\hat{y}_k)\geq L\bar{R}_{k}$, we apply the chain rule to linearize them around $\mathbf{u}^{(j)}$. Since $x_k$ is determined through the dynamics \eqref{eq:controllable_dynamics} and $\hat{y}_k$ is predicted via $\Omega(x_0, y_0, \mathbf{u}_{0:k-2})$, the gradient of $c(x_k,\hat{y}_k)$ with respect to $\mathbf{u}$ is:
\begin{equation}
    \nabla_{\mathbf{u}} c(x_k,\hat{y}_k) = \nabla_{x} c(x_k,\hat{y}_k) \cdot \nabla_{\mathbf{u}} x_k + \nabla_{y} c(x_k,\hat{y}_k) \cdot \nabla_{\mathbf{u}} \hat{y}_k,
\end{equation}
where $\nabla_{\mathbf{u}} x_k$ has been computed as above and $\nabla_{\mathbf{u}} \hat{y}_k$ is obtained from the neural network predictor $\Omega$. The linearized safety constraint around $\mathbf{u}^{(j)}$ is then:
\begin{equation}
    \label{eq:linearized_constraint}
    \tilde{c}_k^{(j)}(\mathbf{u}) = c(x_k^{(j)},\hat{y}_k^{(j)})+\nabla_{\mathbf{u}} c(x_k^{(j)},\hat{y}_k^{(j)})^\top(\mathbf{u}-\mathbf{u}^{(j)}).
\end{equation}

To avoid artificial unboundedness (\cite{Mao_2016}), where the solution of the linearized problem may lie far from the linearization trajectory $\mathbf{u}^{(j)}$, it is necessary to add trust region
constraints $\|\mathbf{u}-\mathbf{u}^{(j)}\|\leq \Delta_j$ where
$\Delta_j\in [0,\Delta_0]$, $\Delta_0 >0$ is the trust region radius. 
To guarantee convergence to a local optimum, the trust region radius $\Delta_j$ is typically updated adaptively based on the ratio of the actual nonlinear cost reduction to the predicted linearized cost reduction. It shrinks only when the linear approximation is poor, preventing the algorithm from premature termination before reaching a KKT point~(\cite{cc-scp})
This leads to the linearized optimal control problem (LOCP) at iteration $j + 1$  defined as follows.




\begin{prob}[\textbf{LOCP}]\upshape
\label{prb:locp}
Find a control input sequence by
$$\min_{\mathbf{u} }  J(\mathbf{x},\mathbf{u})$$
subject to, for all $k = 1, \dots, T$:
\begin{align*}
     & \tilde{c}_k^{(j)}(\mathbf{u}) \geq L \bar{R}_{k},\\
     & u_k \in \mathcal{U}, \tilde{x}_k^{(j)}(\mathbf{u}) \in \mathcal{X},\\
    & \|\mathbf{u}-\mathbf{u}^{(j)}\|\leq \Delta_j. 
\end{align*}
\end{prob}

\section{Further Analysis of Algorithmic Properties} 
\label{sec:convergence}

In the previous section, we presented the algorithm and its basic properties. 
However, two questions still remain:
(i) why is the shortcut solution KKT optimal; and 
(ii) if neither the shortcut nor the infeasible-reject condition is triggered, does the outer loop converge in finitely many iterations to an $\epsilon_{\mathrm{tol}}$-optimal solution? 
This section  establishes the convergence and optimality properties of the proposed two-loop algorithm, directly addressing the two questions posed above. We begin by establishing the theoretical foundation provided by the inner loop.  
The inner SCP loop is guaranteed to converge to a solution that satisfies the Karush--Kuhn--Tucker (KKT) conditions for a specific fixed-bound DOCP~(\cite{cc-scp}). However, updating the error bounds via Eq.~\eqref{eq:construct_bound} dynamically alters the feasible region, formulating a new DOCP. Consequently, a KKT-optimal solution obtained in the $\ell$-th iteration may lose both its feasibility and its optimality in the $(\ell+1)$-th iteration. 

To systematically analyze this, the remainder of the section is structured into two parts. First, we formally prove that under specific mathematical conditions, the algorithm returns a KKT-optimal solution. In such cases, the structural optimality of the current solution is perfectly inherited, mathematically justifying the early-termination \emph{shortcut} (Theorem~\ref{thm:outer_convergence}). Subsequently, we show that in a general scenario where the shortcut and feasibility-reject mechanism are not triggered, the outer loop converges in finitely many iterations to an $\epsilon_{\mathrm{tol}}$-optimal solution.


For notational clarity, we reformulate the $\ell$-th DOCP (Problem~\ref{prb:docp}) in a  general and concise form as: 
\begin{equation}
\label{eq:simply_ocp}
\min_{\mathbf{u}} o^{(\ell)}(\mathbf{u}) \quad \text{s.t.}~ h^{(\ell)}(\mathbf{u})=\mathbf{0},~g^{(\ell)}(\mathbf{u}) \leq \mathbf{0}.
\end{equation}
and write the LOCP (Problem~\ref{prb:locp}) at outer loop iteration $\ell$, inner loop iteration $j+1$  as:
\begin{equation}
\label{eq:simply_locp}
\begin{aligned}
&\min_{\mathbf{u}} o^{(\ell)}(\mathbf{u}^{(\ell,j)})+\nabla_{\mathbf{u}} o^{(\ell)}(\mathbf{u}^{(\ell,j)})(\mathbf{u}-\mathbf{u}^{(\ell,j)}) \\
&\text{s.t.} \quad g^{(\ell)}(\mathbf{u}^{(\ell,j)})+\nabla_{\mathbf{u}} g^{(\ell)}(\mathbf{u}^{(\ell,j)})(\mathbf{u}-\mathbf{u}^{(\ell,j)}) \leq \mathbf{0}
\end{aligned}
\end{equation}
Inner loop convergence relies on the following assumptions, which we briefly outline and discuss their validity in the context of the DOCP. 

\emph{Notation.} Throughout the convergence analysis of the inner loop, we work within a single outer iteration $\ell$.
For notational simplicity, we drop the outer index in  \eqref{eq:simply_locp} and write $o^{(\ell)}\to o$, $g^{(\ell)}\to g$, and $\mathbf{u}^{(\ell,j)}\to \mathbf{u}^{(j)}$.

\begin{assum}
\label{as:1}
    The functions $o(\mathbf{u})$ and $g(\mathbf{u})$ are both $C^1$ with respect to $\mathbf{u}$. 
\end{assum}

\begin{assum}
\label{as:2}
At each iteration $j$, \eqref{eq:simply_locp} has a solution $\mathbf{u}^{(j)}$. Moreover, $\mathbf{u}^{(j)}$ satisfies the Linear Independence Constraint Qualification(LICQ) related to \eqref{eq:simply_locp}. Finally, the family of solutions $\{\mathbf{u}^{(j)}\}_{j \in \mathbb{N}}$ is bounded.
\end{assum}

\begin{assum}
    \label{as:3}
    Define $\Delta \mathbf{u}^{(j+1)}:=\mathbf{u}^{(j+1)}-\mathbf{u}^{(j)}$, we require that there exists
    $ \mathcal{J}>0, \text{s.t.}, \forall{j>\mathcal{J}}, \|\Delta \mathbf{u}^{(j+1)}\|<\|\Delta \mathbf{u}^{(j)}\|.$
\end{assum}

Assumption~\ref{as:1} ensures that the first-order Taylor approximations used in the SCP iterations are well-defined. 
For the functions $o$ and $g$ to be continuously differentiable, we require: (a) the safety function $c$ to be $C^1$ in both $x$ and $y$; (b) the dynamic function $f$ to be $C^1$ in both $u$ and $x$ 
; and (c) the neural network predictor $\Omega$ to be $C^1$, which requires the nonlinear layers to use differentiable activation functions such as sigmoid. 
Assumption~\ref{as:2} is classic in convex optimization and easily satisfied in the context of the linearized problems. 
Assumption~\ref{as:3} is satisfied due to the trust region constraint introduced in Problem~\ref{prb:locp}. It is important to note that Assumption~\ref{as:3} does not require the entire sequence $\{\mathbf{u}^{(j)}\}_{j \in \mathbb{N}}$ to converge.

\begin{lem}(Convergence guarantees, \cite{cc-scp}). 
\label{thm:convergence_inner}
Assume that Assumptions~\ref{as:1},\ref{as:2},\ref{as:3} hold and consider the family $\{\mathbf{u}^{(j)}\}_{j \in \mathbb{N}}$ where $\mathbf{u}^{(j)}$ is solution of \eqref{eq:simply_locp} at iteration $j$. The following holds:
\begin{enumerate}
    \item If there exists an iteration $\bar{j}$ such that for every $j \geq \bar{j}$ it holds $\mathbf{u}^{(j)}=\mathbf{u}^{(\bar{j})}$, then $\mathbf{u}^{(\bar{j})}$ is a feasible point satisfying the KKT conditions related to \eqref{eq:simply_ocp}.
    \item Assume that $\{\mathbf{u}^{(j)}\}_{j\in \mathbb{N}}$ is an infinite sequence of solution for \eqref{eq:simply_locp}. Then, there exists a subsequence that converges to a point $\overline{\mathbf{u}}$ satisfying the KKT conditions related to \eqref{eq:simply_ocp}.
\end{enumerate}
In both cases, denote by $\mathbf{u}^*$ the resulting point; that is, set $\mathbf{u}^* := \mathbf{u}^{(\bar{j})}$ in the finite-termination case and $\mathbf{u}^* := \overline{\mathbf{u}}$ in the infinite-sequence case.
\end{lem}

Having established the convergence of the inner SCP loop for fixed error bounds, we now turn to the outer loop, which iteratively updates the error bounds $\bar{R}^{(\ell)}$ and re-solves the DOCP. The key question is whether the KKT solution from one iteration remains valid when the error bounds are changed. Our following theorem provides conditions under which the solution quality is preserved across outer iterations.

\begin{thm}
\label{thm:outer_convergence}
Let $\mathbf{u}^{*(\ell)}$ be a KKT solution to the $\ell$-th DOCP
obtained by the SCP (Lemma.~\ref{thm:convergence_inner}). Using $\mathbf{u}^{*(\ell)}$, construct the updated error bounds $\bar{R}^{(\ell+1)}$ and hence the constraints $g^{(\ell+1)}$ for the ($\ell$+1)-th DOCP. If it is verified that $g^{(\ell+1)}(\mathbf{u^{*(\ell)}})\leq 0$, 
and the bounds are non-decreasing, i.e., $\{\bar{R}_k^{(\ell+1)}\geq\bar{R}_k^{(\ell)}\}_{k=1}^T$, then $\mathbf{u}^{*(\ell)}$ satisfies the KKT conditions for the $(\ell+1)$-th DOCP with the same multiplier $(\alpha, \lambda, \zeta)$.

\end{thm}

 \begin{pf}By Lemma~\ref{thm:convergence_inner}, $\mathbf{u}^{*(\ell)}$ is a feasible point satisfying the KKT conditions for \eqref{eq:simply_ocp} with error bounds $\bar{R}^{(\ell)}$. 

Let us denote the objective and constraint functions at the two DOCP by $(o^{(\ell)}, h^{(\ell)}, g^{(\ell)})$ and $(o^{(\ell+1)}, h^{(\ell+1)}, g^{(\ell+1)})$, respectively. Note that the inequality constraint vector $g$ consists of two parts: $g = [g^{\text{state}}, g^{\text{safe}}]$, where $g^{\text{state}}$ represents the state constraints $\tilde{x}_k^{(j)}(\mathbf{u}) \in \mathcal{X}$ and $g^{\text{safe}} = \{L\bar{R}_k - c(x_k,\hat{y}_k)\}_{k=1}^T$ represents the safety constraints. Similarly, the multiplier $\zeta$ can be partitioned as $\zeta = [\zeta^{\text{state}}, \zeta^{\text{safe}}]$.

Since only the error bounds change between iterations, we have $o^{(\ell)}=o^{(\ell+1)}$, $h^{(\ell)}=h^{(\ell+1)}$, and $g^{\text{state},(\ell)} = g^{\text{state},(\ell+1)}$. However, for the safety constraints,
$$g^{\text{safe},(\ell+1)}(\mathbf{u})-g^{\text{safe},(\ell)}(\mathbf{u})=L\big(\bar{R}^{(\ell+1)}-\bar{R}^{(\ell)}\big).$$
If $\{\bar{R}_k^{(\ell+1)}=\bar{R}_k^{(\ell)}\}_{k=1}^T$, the statement holds trivially. We therefore focus on the case where $\{\bar{R}_k^{(\ell+1)}>\bar{R}_k^{(\ell)}\}_{k=1}^T$. We verify the KKT conditions for $\mathbf{u}^{*(\ell)}$ with respect to \eqref{eq:simply_ocp} using $\bar{R}^{(\ell+1)}$:
\begin{enumerate}
    \item \textbf{Stationarity:} Since $\mathbf{u}^{*(\ell)}$ satisfies the KKT conditions for the previous iteration, we have:
    \begin{align*}
        &\alpha \, \nabla_{\mathbf{u}} o^{(\ell)}(\mathbf{u}^{*(\ell)})+ {\lambda}^{\top} \, \nabla_{\mathbf{u}}  h^{(\ell)}(\mathbf{u}^{*(\ell)})\\
        &+(\zeta^{\text{state}})^{\top} \, \nabla_{\mathbf{u}}  g^{\text{state},(\ell)}(\mathbf{u}^{*(\ell)}) \\
        &+ (\zeta^{\text{safe}})^{\top} \, \nabla_{\mathbf{u}}  g^{\text{safe},(\ell)}(\mathbf{u}^{*(\ell)})=\mathbf{0}.
    \end{align*}
    Since $o^{(\ell)}=o^{(\ell+1)}$, $h^{(\ell)}=h^{(\ell+1)}$, $g^{\text{state},(\ell)} = g^{\text{state},(\ell+1)}$, and $\bar{R}$ enters $g^{\text{safe}}$ only as a constant offset independent of $\mathbf{u}$, we have $\nabla_{\mathbf{u}} g^{\text{safe},(\ell)}(\mathbf{u}) = \nabla_{\mathbf{u}} g^{\text{safe},(\ell+1)}(\mathbf{u})$ for all $\mathbf{u}$. It follows that
    \begin{align*}
        &\alpha \, \nabla_{\mathbf{u}} o^{(\ell+1)}(\mathbf{u}^{*(\ell)})+ {\lambda}^{\top} \, \nabla_{\mathbf{u}}  h^{(\ell+1)}(\mathbf{u}^{*(\ell)})\\
        &+ {\zeta}^{\top} \, \nabla_{\mathbf{u}}  g^{(\ell+1)}(\mathbf{u}^{*(\ell)})=\mathbf{0}.
    \end{align*}
    
    \item \textbf{Primal feasibility:} By the theorem's hypothesis, $g^{(\ell+1)}(\mathbf{u}^{*(\ell)})\leq \mathbf{0}$.
    
    \item \textbf{Complementary slackness:} We need to prove that $\zeta_s \cdot g_s^{(\ell+1)}(\mathbf{u}^{*(\ell)}) = 0$ for all $s \in \{1,\ldots,2T\}$. 
    
    For state constraint indices $s \in \{1,\ldots,T\}$: Since $g^{\text{state},(\ell+1)}_s = g^{\text{state},(\ell)}_s$, 
    we have $\zeta^{\text{state}}_s \cdot g^{\text{state},(\ell+1)}_s(\mathbf{u}^{*(\ell)}) = \zeta^{\text{state}}_s \cdot g^{\text{state},(\ell)}_s(\mathbf{u}^{*(\ell)}) = 0$.
    
    For safety constraint indices $s \in \{T+1,\ldots,2T\}$: We show this via contradiction. Assume $\exists s$ such that $\zeta^{\text{safe}}_s > 0$. From the KKT condition at iteration $\ell$, we must have $g^{\text{safe},(\ell)}_s(\mathbf{u}^{*(\ell)}) = 0$, as $\zeta^{\text{safe}}_s > 0$ implies active constraint. 
    Since $\forall k=s-T,\bar{R}_k^{(\ell+1)} > \bar{R}_k^{(\ell)}$ and $L > 0$, we have $g^{\text{safe},(\ell+1)}_s(\mathbf{u}^{*(\ell)}) = g^{\text{safe},(\ell)}_s(\mathbf{u}^{*(\ell)}) + L(\bar{R}_k^{(\ell+1)} - \bar{R}_k^{(\ell)}) > 0$, contradicting the feasibility condition $g^{\text{safe},(\ell+1)}_s(\mathbf{u}^{*(\ell)}) \leq 0$.
    Hence, our assumption must be false, and we conclude that $\forall s\in\{T+1,\ldots,2T\},\zeta^{\text{safe}}_s = 0$, thus $\zeta^{\text{safe}}_s \cdot g^{\text{safe},(\ell+1)}_s(\mathbf{u}^{*(\ell)}) = 0$.\qed
    
\end{enumerate}
\end{pf}


The above results establish KKT optimality under the shortcut condition. Geometrically, when the error bounds monotonically non-decrease ($\bar{R}^{(\ell+1)} \ge \bar{R}^{(\ell)}$), the required safety margins expand, causing the feasible set of the optimization problem to shrink. If the previous optimizer $\mathbf{u}^{*(\ell)}$ remains feasible within this contracted set, it must remain optimal. Because the tightening of the safety constraints does not alter the local gradient landscape or the active constraint set at $\mathbf{u}^{*(\ell)}$, its structural optimality is perfectly preserved without needing further SCP iterations.

Next, we discuss the convergence behavior of the outer loop when neither the shortcut nor the infeasible-reject condition is triggered beforehand. In this normal iterative scenario, the algorithm guarantees a monotonically decreasing cost sequence $\{J(\mathbf{u}^{*(\ell)})\}$. This monotonic descent is structurally enforced by the infeasibility-reject mechanism (Algorithm~\ref{alg:scp_outer}, Line 8): the algorithm proceeds to the $(\ell+1)$-th iteration only if the current KKT solution $\mathbf{u}^{*(\ell)}$ remains strictly feasible under the newly constructed error bounds $\bar{R}^{(\ell+1)}$. Because the subsequent inner loop SCP is a descent method initialized at this already-feasible point $\mathbf{u}^{*(\ell)}$, the optimization strictly guarantees that $J(\mathbf{u}^{*(\ell+1)}) \le J(\mathbf{u}^{*(\ell)})$.

Furthermore, this monotonic descent ensures convergence in finitely many steps. Because the input space is compact and $J$ is continuous, the cost function is bounded below by a finite minimum $J_{\min}$. For normal outer iterations that do not trigger the shortcut, the cost must strictly decrease by an amount greater than the tolerance, i.e., $J(\mathbf{u}^{*(\ell-1)}) - J(\mathbf{u}^{*(\ell)}) > \epsilon_{\mathrm{tol}}$. Since the maximum possible reduction in the cost function is strictly finite ($J(\mathbf{u}^{*(0)}) - J_{\min} < \infty$), the total number of such outer iterations is universally upper-bounded by $(J(\mathbf{u}^{*(0)}) - J_{\min}) / \epsilon_{\mathrm{tol}}$. 
When the sequence converges and terminates with $J(\mathbf{u}^{*(\ell-1)}) - J(\mathbf{u}^{*(\ell)}) \le \epsilon_{\mathrm{tol}}$, the algorithm returns $\mathbf{u}^{*(\ell-1)}$. It is strictly feasible with its own corresponding error bounds. and rigorously bounded as an $\epsilon_{\mathrm{tol}}$-optimal solution around the exact KKT optimal solution for that specific feasible region.

\section{Simulation and Results}
In this section, we demonstrate and evaluate the proposed approach through simulations of an autonomous driving case study involving uncontrollable pedestrians.
Specifically, we develop a high-fidelity simulator, where the behavior of pedestrians, influenced by the vehicle, is modeled using a multi-parameter Social Force Model (\cite{sfm}). It has been validated against a large set of real-world data (\cite{citr}).
Our entire framework is implemented in Python 3, and all codes are available on our project website.\footnote{\url{https://github.com/Yangming911/Conformal_Tube_MPC}}

\subsection{Scenario Description and System Dynamics}
\begin{figure}[t]
  \centering
  \includegraphics[width=0.44\textwidth]{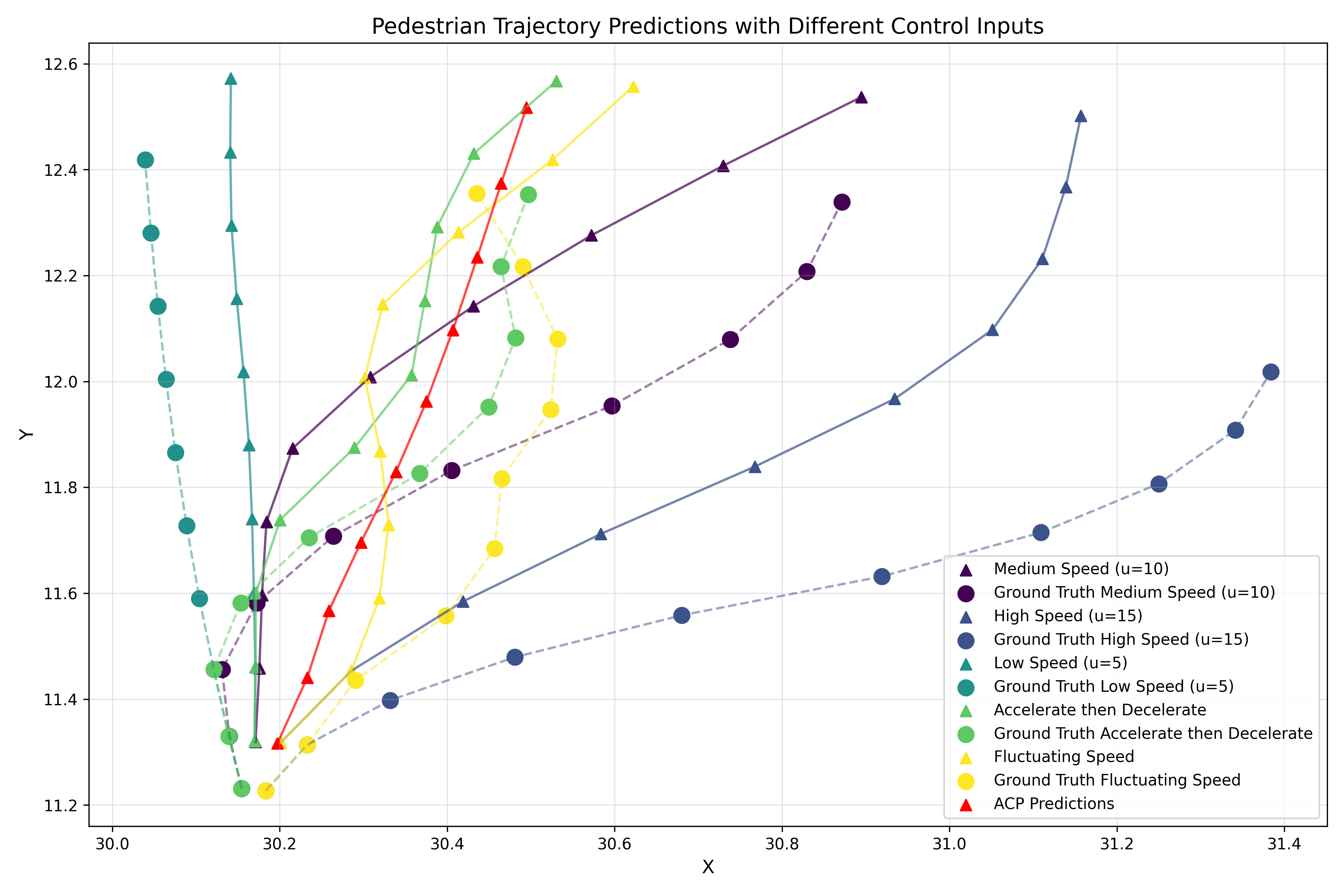}
  \caption{Scenario Illustration: given the same initial position, the distribution of a pedestrian's trajectory over the next few steps will change with the car's control signals. Dashed lines with circular markers in different colors illustrate varying pedestrian behavioral patterns under different vehicle control policies. Solid lines with triangular markers of the same color represent the corresponding neural network predictions. Because pedestrian behavior is inherently stochastic, point predictions can never be perfectly accurate, necessitating reliable error bound coverage for safety-critical control.}
  \label{fig:traj_vs_control}
\end{figure}

\textbf{Scenario Setup:} A straight road segment with a total length of 50 m. A crosswalk is located within this segment, and the road has a width of 20 m. The starting line for the vehicle is positioned at one end of the road, while the ending line is at the opposite end. Several pedestrians are present near the crosswalk, and a vehicle is approaching from the starting line. The control period is set to be $\Delta t $ = 0.1s. The car needs to pass the intersection as fast as possible while keeping a safe distance from pedestrians. 

\textbf{Vehicle Dynamics:}
To demonstrate the generalizability of our algorithm, we consider two types of vehicle dynamics modeling.

In the linear model, we assume the vehicle moves straight through a crosswalk, and its state is given by 
\[
\mathbf{x} = \mathbf{x}_0 + \mathbf{L}\mathbf{u},
\]
where $\mathbf{L} \in \mathbb{R}^{T \times T}$ is a lower triangular matrix with $l_{ij} = 1$ for $i \geq j$ and $l_{ij} = 0$ otherwise. Here, $\mathbf{x} \in \mathbb{R}^{T}$ and $\mathbf{u} \in \mathbb{R}^{T}$ denote the position and control input (speed) sequences over the horizon, respectively. This assumption aligns with real-world constraints, as vehicles are generally not permitted to maneuver left or right while crossing a zebra crossing.

In the nonlinear model, we further incorporate the effect of wind resistance. The state evolution is formulated as
\[
\mathbf{x} = \mathbf{x}_0 + \mathbf{L} \left( \mathbf{u} - \frac{1}{2}\sigma_{\text{wind}} (\mathbf{u} + v_{\text{wind}}\mathbf{1})^{\circ 2} \right),
\]
where $v_{\text{wind}}$ denotes the wind speed (set to 0.5), $\sigma_{\text{wind}}$ is the wind resistance coefficient (set to 0.01), $\mathbf{1}$ is an all-ones vector of compatible dimension, and $(\cdot)^{\circ 2}$ represents element-wise squaring.

Nevertheless, our approach naturally extends to two or higher dimensions. The vehicle speed is constrained within $u_t \in [0, 15]$ m/s, reflecting typical urban speed limits.

\textbf{Pedestrian Behavior:}
The pedestrian movement is simulated using the \emph{Social Force Model} introduced by \cite{sfm}, which treats pedestrians as self-driven particles influenced by multiple force components that govern their dynamic behavior. The pedestrian dynamics are governed by a force-based equation $m \frac{d\vec{v}(t)}{dt} = \vec{F}_{\text{vehicle}}(t) + \vec{F}_{\text{destination}}(t) + \vec{\xi}(t)$.

Note that both $\vec{F}_{\text{vehicle}}$ and $\vec{F}_{\text{destination}}$ are complicated nonlinear functions involving parameters shown in \cite{sfm}. 
Here we would like to remark that this model is only used for the purpose of data collection and validation; its structure and parameters are unknown to the control designer.

\textbf{Data Collection and Network Predictor:}  
The size of the training dataset $D_{\text {train }}$ is $2 \times 10^5$, and the size of the calibration dataset $D_{\text {cal}}$ is $2 \times 10^5$. 
We evaluate the controller over 200 trajectories.
We implemented a Seq2Seq neural network for pedestrian position prediction using recurrent architectures with residual connections. The architecture employs a 2-layer GRU (Gated Recurrent Unit, \cite{chung2014empiricalevaluationgatedrecurrent}) with 128 hidden units per layer. The network incorporates layer normalization and residual connections through cumulative delta predictions\footnote{Training utilizes the Adam optimizer, an MSE loss function computed over full sequences, and the PyTorch ReduceLROnPlateau scheduler, on an NVIDIA GeForce RTX 3090.}.

\textbf{Safety Constraint:} For vehicle-pedestrian collision avoidance, we frame a Lipschitz continuous function $c(x,y) =  ||x - y||_2 - d_\text{safe}$,
where $d_{\text{safe}}$ = 2.0 m.
 

\begin{table*}[t]
\centering
\caption{Success rate, average speed, acceleration, solving time, and PDM score for our SCP$^2$ compared to other methods under linear system dynamics scenario and M uncontrollable agents}.
\begin{adjustbox}{max width=\textwidth}
\begin{tabular}{|l|ccc|ccc|ccc|ccc|ccc|}
\hline
                           & \multicolumn{3}{c|}{Success Rate}                                             & \multicolumn{3}{c|}{Average Speed {[}$m/s${]}}                      & \multicolumn{3}{c|}{Average Acc.
                           {[}$m/s^2${]}}             &\multicolumn{3}{c|}{Average Solv. Time {[}$ms${]}}  & \multicolumn{3}{c|}{PDM Score}                \\ \hline
\diagbox[innerwidth=0.8cm,height=0.5cm]{Alg.}{\raisebox{-0.5ex}{M}}  & \multicolumn{1}{c|}{1}       & \multicolumn{1}{c|}{5}      & 9               & \multicolumn{1}{c|}{1}     & \multicolumn{1}{c|}{5}    & 9   & \multicolumn{1}{c|}{1}     & \multicolumn{1}{c|}{5}    & 9   & \multicolumn{1}{c|}{1}       & \multicolumn{1}{c|}{5}       & 9  & \multicolumn{1}{c|}{1}       & \multicolumn{1}{c|}{5}       & 9                \\ \hline

APF                & \multicolumn{1}{c|}{88.5\%} & \multicolumn{1}{c|}{68.0\%} & 70.0\%          & \multicolumn{1}{c|}{2.83} & \multicolumn{1}{c|}{4.50} & 3.62 & \multicolumn{1}{c|}{\textbf{0.09}} & \multicolumn{1}{c|}{\textbf{0.17}} & \textbf{0.15}   & \multicolumn{1}{c|}{\textbf{0.01}} & \multicolumn{1}{c|}{\textbf{0.02}} & \textbf{0.03}  & \multicolumn{1}{c|}{77.68} & \multicolumn{1}{c|}{62.39} & 63.41     \\ \hline
ACP          & \multicolumn{1}{c|}{97.0\%} & \multicolumn{1}{c|}{92.5\%} & 93.5\%          & \multicolumn{1}{c|}{9.64} & \multicolumn{1}{c|}{7.04} & 6.40 & \multicolumn{1}{c|}{0.98} & \multicolumn{1}{c|}{3.03} & 4.53  & \multicolumn{1}{c|}{22.40} & \multicolumn{1}{c|}{36.28} & 40.42  & \multicolumn{1}{c|}{88.99} & \multicolumn{1}{c|}{83.59} & 83.92       \\ \hline
SCP$^2$                & \multicolumn{1}{c|}{\textbf{99.5}\%} & \multicolumn{1}{c|}{\textbf{99.5}\%} & \textbf{100.0}\%          & \multicolumn{1}{c|}{\textbf{12.52}} & \multicolumn{1}{c|}{\textbf{8.39}} & \textbf{7.43} & \multicolumn{1}{c|}{0.77} & \multicolumn{1}{c|}{1.34} & 1.28  & \multicolumn{1}{c|}{105.06} & \multicolumn{1}{c|}{110.13} & 117.24  & \multicolumn{1}{c|}{\textbf{92.92}} & \multicolumn{1}{c|}{\textbf{90.15}} & \textbf{89.91}        \\ \hline

\end{tabular}
\label{table:result}
\end{adjustbox}
\end{table*}

\begin{table*}[t]
\centering
\caption{Our method compared to other methods under the nonlinear system dynamics scenario.}
\begin{adjustbox}{max width=\textwidth}
\begin{tabular}{|l|ccc|ccc|ccc|ccc|ccc|}
\hline
                           & \multicolumn{3}{c|}{Success Rate}                                             & \multicolumn{3}{c|}{Average Speed {[}$m/s${]}}                      & \multicolumn{3}{c|}{Average Acc.                           {[}$m/s^2${]}}       &\multicolumn{3}{c|}{Average Solv. Time {[}$ms${]}}        & \multicolumn{3}{c|}{PDM Score}                \\ \hline
\diagbox[innerwidth=0.8cm,height=0.5cm]{Alg.}{\raisebox{-0.5ex}{M}}  & \multicolumn{1}{c|}{1}       & \multicolumn{1}{c|}{5}      & 9               & \multicolumn{1}{c|}{1}     & \multicolumn{1}{c|}{5}    & 9    & \multicolumn{1}{c|}{1}       & \multicolumn{1}{c|}{5}       & 9  & \multicolumn{1}{c|}{1}     & \multicolumn{1}{c|}{5}    & 9   & \multicolumn{1}{c|}{1}       & \multicolumn{1}{c|}{5}       & 9                \\ \hline

APF                & \multicolumn{1}{c|}{97.5\%} & \multicolumn{1}{c|}{88.5\%} & 84.5\%          & \multicolumn{1}{c|}{2.01} & \multicolumn{1}{c|}{2.43} & 2.61 & \multicolumn{1}{c|}{\textbf{0.05}} & \multicolumn{1}{c|}{\textbf{0.06}} & \textbf{0.07}   & \multicolumn{1}{c|}{\textbf{0.01}} & \multicolumn{1}{c|}{\textbf{0.01}} & \textbf{0.02}  & \multicolumn{1}{c|}{85.11} & \multicolumn{1}{c|}{79.20} & 76.55     \\ \hline
ACP,           & \multicolumn{1}{c|}{100.0\%} & \multicolumn{1}{c|}{100.0\%} & 100.0\%          & \multicolumn{1}{c|}{9.86} & \multicolumn{1}{c|}{7.46} & 6.90 & \multicolumn{1}{c|}{1.57} & \multicolumn{1}{c|}{4.65} & 4.62   & \multicolumn{1}{c|}{13.96} & \multicolumn{1}{c|}{24.08} & 31.89 & \multicolumn{1}{c|}{90.15} & \multicolumn{1}{c|}{78.51} & 78.04       \\ \hline
SCP$^2$                 & \multicolumn{1}{c|}{100.0\%} & \multicolumn{1}{c|}{100.0\%} & 100.0\%          & \multicolumn{1}{c|}{\textbf{12.87}} & \multicolumn{1}{c|}{\textbf{9.28}} & \textbf{8.45} & \multicolumn{1}{c|}{0.60} & \multicolumn{1}{c|}{1.59} & 1.67   & \multicolumn{1}{c|}{98.62} & \multicolumn{1}{c|}{100.99} & 103.88 & \multicolumn{1}{c|}{\textbf{96.07}} & \multicolumn{1}{c|}{\textbf{89.51}} & \textbf{88.44}        \\ \hline

\end{tabular}
\label{table:nonlinear_result}
\end{adjustbox}
\end{table*}

\begin{figure}[t]
  \centering
  \includegraphics[width=0.48\textwidth]{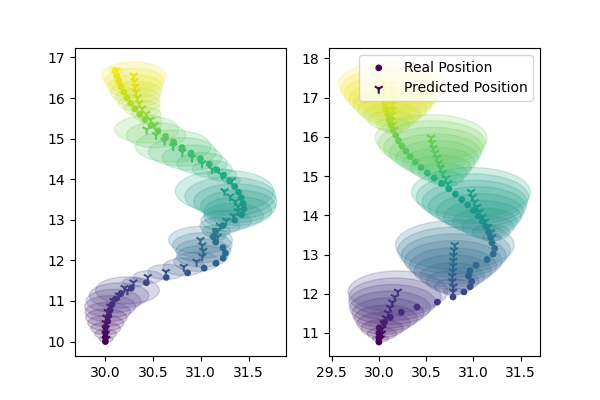}
\caption{Pedestrian trajectory ground truth, and network predictions with error bounds under SCP$^2$ (left) and ACP (right). Our proposed SCP$^2$ generates tighter, region-specific error bounds that rapidly adapt to the control policy and the pedestrian's nonlinear maneuvers. In contrast, the baseline ACP produces overly conservative bounds and struggles to capture the trajectory curvature efficiently due to its time-lagged adaptation mechanism.}
  \label{fig:prediction_compare}
\end{figure}

\begin{figure}[t]
  \centering
  \includegraphics[width=0.48\textwidth]{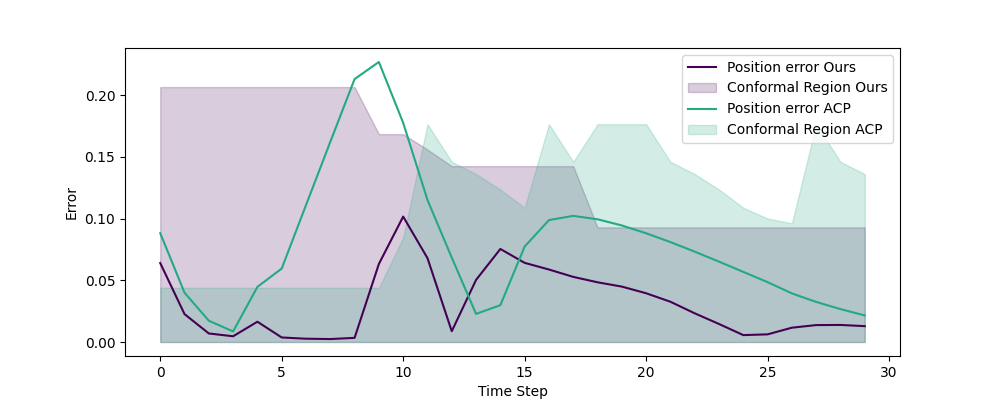}
  \caption{Prediction Error with Bound $\bar{R}_{k=1}$ through a complete zebra crossing task (SCP$^2$ v.s. ACP). SCP$^2$ can correctly predict error bounds (85\% coverage) even under rapidly changing multi-agent interaction. The lines are the true prediction errors, and the areas are the error bounds.} 
  \label{fig:error_compare}
\end{figure}

\subsection{Results}
We evaluate multi-pedestrian navigation under varying crowd densities using a simulator with $M\in\{1,5,9\}$ pedestrians following social-force dynamics. We compare a classic artificial potential field (APF, \cite{apf}), Adaptive Conformal Prediction (ACP, \cite{dixit2023adaptive}), and our Robust Split Conformal Prediction-Sequential Convex Programming (SCP$^2$). SCP$^2$ and ACP have the same failure probability tolerance ($\alpha=0.15$) and MPC time horizon ($T=10$). Note that ACP requires the optimal problem to be convex. In our non-convex collision avoidance scenario, we use the non-convex optimization solver SLSQP in SciPy to implement it. We tested the above algorithm on both linear and nonlinear dynamic models mentioned above.

The experimental results are presented in Table \ref{table:result} and Table \ref{table:nonlinear_result}. The performance is evaluated based on four metrics.
The PDM score introduced in \cite{Dauner2024ARXIV} is widely used in the evaluation of autonomous driving tasks, computed by:
\begin{equation}
        \text{PDM Score} = \epsilon_1 \cdot \text{SR}_{norm} + \epsilon_2 \cdot \text{V}_{norm}+ \epsilon_3 \cdot \text{C}_{norm},
\end{equation}
$\text{SR}_{norm},\text{V}_{norm},\text{C}_{norm}$ represent normalized success rate, average speed, and comfort level, respectively, with comfort being inversely proportional to average acceleration\footnote{We assigned weights \(\epsilon_1\), \(\epsilon_2\), \(\epsilon_3\) = 0.8,0.1,0.1 to each item, since safety is our primary concern in our experimental scenarios. }.
The results clearly demonstrate that the proposed SCP$^2$ outperforms both the APF and ACP algorithms across nearly all evaluated metrics, including the overall PDM Score, in both the linear and nonlinear models. It achieves the highest success rate, operates at higher speeds, and maintains smooth acceleration under both system settings.

To further illustrate the effectiveness of our approach, Fig.~\ref{fig:prediction_compare} visualizes a representative zebra-crossing scenario where the vehicle navigates around a pedestrian. 
Both methods successfully predict the pedestrian's future positions and construct error bounds (shown with color gradients indicating time progression). Notably, SCP$^2$ produces tighter and more rapidly adaptive error bounds that closely follow the actual pedestrian trajectory, enabling the vehicle to plan more efficient paths while maintaining safety guarantees. 
Fig.~\ref{fig:error_compare} illustrates that SCP$^2$'s bounds expand appropriately to capture increased uncertainty during critical interaction phases, then contract during more predictable periods. In contrast, ACP exhibits several limitations. First, its error bounds remain unchanged during the initial $T$ time steps due to the time-lagged nature of its evaluation,
which requires a warm-up period before adaptation can begin. Second, ACP adapts slowly to distribution changes because of its time-lagging design to address distribution shifts. It operates in an a posteriori manner—updating bounds based on empirical errors realized from preceding horizons. In contrast, our approach aims for \textit{a priori} robustness to ensure instantaneous safety.
Consequently, the vehicle controlled by our method can smoothly anticipate pedestrian behaviors. It safely navigates with proactive, gentle adjustments rather than abrupt, reactive braking, producing a smoother acceleration profile and improving speed efficiency. However, this anticipatory capability comes at the cost of significantly increased problem complexity. Addressing this requires our proposed two-loop architecture, which incurs a higher computational burden. Nevertheless, the resulting computation time—on the order of 100 ms—fully satisfies the operational requirements for a high-level behavioral and motion planner, as opposed to a low-level dynamic tracking controller.

The early termination mechanism described in Theorem~\ref{thm:outer_convergence} significantly reduces computational cost. Table~\ref{table:shortcut_comparison} demonstrates the effectiveness of this shortcut mechanism across different scenarios.
\begin{table}[t]
\caption{Impact of Early Termination Shortcut on Computational Efficiency}
\label{table:shortcut_comparison}
\centering
\begin{tabular}{|c|c| c|}
\hline
$M$ & \begin{tabular}[c]{@{}c@{}}\\Time (ms)\\ w/o $\rightarrow$ w/ shortcut\end{tabular} & \begin{tabular}[c]{@{}c@{}}Avg. Outer Loop Iter.\\ w/o $\rightarrow$ w/ shortcut\end{tabular} \\ \hline
1 & $204.60 \rightarrow 189.17$ & $4.36 \rightarrow 2.63$ \\ \hline
5 & $212.21 \rightarrow 202.59$ & $3.33 \rightarrow 2.23$ \\ \hline
9 & $236.69 \rightarrow 225.73$ & $3.09 \rightarrow 2.16$ \\ \hline
\end{tabular}
\end{table}
The results show that the shortcut mechanism consistently reduces the average outer loop iterations per MPC solve by approximately 30-40\% across all scenarios, and reduces the total computation time by 7- 12

\subsection{Analysis of Robust Conformal Prediction}
\begin{figure}[t]
  \centering
  \includegraphics[width=0.48\textwidth]{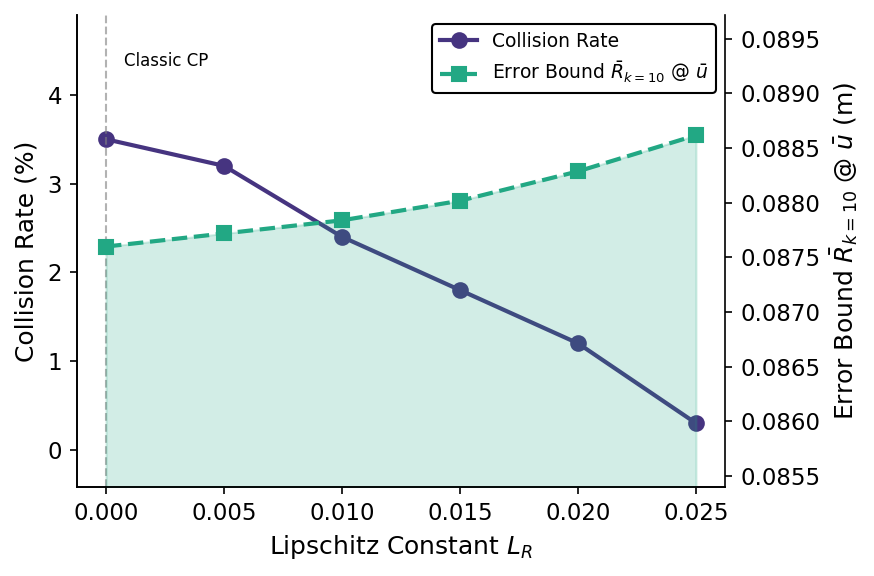}
  \caption{Collision rate and error bound $\bar{R}_{k=10}$ at average speed with varying Lipschitz constant $L_R$. Setting $L_R=0$ corresponds to classic CP.}
  \label{fig:robust_cp_exp1}
\end{figure}

\begin{figure}[t]
  \centering
  \includegraphics[width=0.48\textwidth]{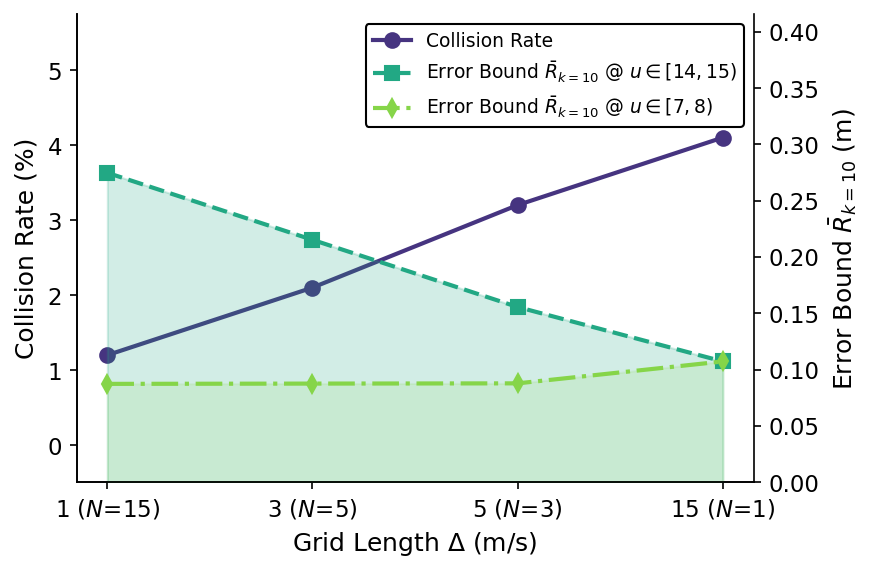}
  \caption{Collision rate and error bound $\bar{R}_{k=10}$ at high/medium speed ranges, with varying grid length $\Delta$ (and corresponding number of grid cells $N$). Notably, as the partition becomes coarser (larger $\Delta$), the high-speed error bound is artificially diluted by the dominant low-error data, leading to under-approximated safety margins and a corresponding rise in the collision rate.}
  \label{fig:robust_cp_exp2}
\end{figure}

We conduct ablation studies on the key parameters governing the approximation of distribution shift: the estimated Lipschitz constant $L_R$, and the max partition grid length $\Delta$. The results are summarized in Fig.~\ref{fig:robust_cp_exp1} and Fig.~\ref{fig:robust_cp_exp2}.

As illustrated in Fig.~\ref{fig:robust_cp_exp1}, we fix the grid length and vary the estimated $L_R$. When $L_R = 0$, the method degenerates to the classic split conformal prediction, which yields a smaller error bound $\bar{R}$ but suffers from a higher collision rate (around $3.5\%$) due to the ignorance of distribution shift. As we systematically increase $L_R$, the robustified error bound expands monotonically. This leads to more conservative planning behaviors and a corresponding decrease in the collision rate, eventually satisfying the strict safety ($0.3\%$ collision rate).

Similarly, Fig.~\ref{fig:robust_cp_exp2} demonstrates the effect of the spatial partition size $\Delta$. In practice, a larger $\Delta$ causes the nonconformity scores to be calculated over a broader, less specific region. As observed in the figure, when the grid length increases (e.g., approaching a single unpartitioned space where $N=1$), the calibration data from highly dynamic, high-uncertainty states (high speed, $u \in [14, 15)$) becomes blended with the more abundant, benign data from lower speeds ($u \in [7, 8)$). Consequently, the empirical quantile for high-speed scenarios is artificially diluted by this mixture, causing the resulting error bound $\bar{R}$ to drop significantly. Because this generalized bound fails to capture the specific local worst-case uncertainty of high-speed maneuvers, the MPC planner utilizes an insufficiently tight safety margin, which directly results in a monotonically increasing collision rate.


\section{Conclusion}
We presented a safe control framework for systems interacting with uncontrollable agents whose behaviors are coupled with the ego system's actions. Our approach leveraged the robust conformal prediction technique to formulate the chance-constrained optimization problem as a deterministic one. An iterative sequential convex programming approach was then proposed to effectively solve the problem. 
In contrast to existing conformal prediction approaches that treat prediction models as black boxes independent of control decisions, our framework explicitly accounts for the coupling between control inputs and predicted agent states, achieving tighter and more adaptive uncertainty quantification. Experimental validation in pedestrian-vehicle interaction scenarios was provided to demonstrate the effectiveness of our approach.

\bibliography{ref}

\end{document}